\documentclass[aps,prl,twocolumn,floatfix,reprint]{revtex4-1}
\usepackage[british]{babel}
\usepackage[utf8]{inputenc}
\usepackage{lmodern}
\usepackage[T1]{fontenc}
\usepackage{hyperref}

\usepackage{mathtools} 
\usepackage{amssymb} 
\usepackage{bbm} 

\usepackage[dvipsnames]{xcolor}
\usepackage{tikz}
	\usetikzlibrary{decorations.pathmorphing} 

\usepackage{booktabs} 
\usepackage{tabularx} 
\usepackage{multirow} 

\DeclareMathOperator{\E}{e}
\newcommand{\I}{\mathrm{i}}

\DeclarePairedDelimiter{\bra}{\langle}{\rvert}
\DeclarePairedDelimiter{\ket}{\lvert}{\rangle}
\DeclarePairedDelimiterX{\braket}[2]{\langle}{\rangle}{#1\vert#2}

\DeclareMathOperator*{\ordprod}{\prod\limits^{\vbox to -.25ex{\kern-.25ex\hbox{$\leftharpoonup$}\vss}}}
\DeclareMathOperator*{\ordprodopp}{\prod\limits^{\vbox to -.25ex{\kern-.25ex\hbox{$\rightharpoonup$}\vss}}}

\newcommand{\To}{\cdots\mspace{-1mu}}

\DeclareMathOperator{\id}{\mathbbm{1}} 

\newcommand{\Uqaff}{$U_q(\widehat{\vphantom{t}\smash{\mathfrak{sl}_2}})$}

\begin{document}
		
\title{Resurrecting the partially isotropic Haldane--Shastry model}

\author{Jules Lamers}
\email{julesl@chalmers.se}
\affiliation{Department of Mathematical Sciences, Chalmers University of Technology and University of Gothenburg, SE-412 96 Gothenburg, Sweden}
	
\begin{abstract}
\noindent We present a new and simpler expression for the Hamiltonian of the partially isotropic (\textsc{xxz}-like) version of the Haldane--Shastry model, which was derived by D.~Uglov over two decades ago in an apparently little-known preprint. While resembling the pairwise long-range form of the Haldane--Shastry model our formula accounts for the multi-spin interactions obtained by Uglov. Our expression is physically meaningful, makes hermiticity manifest, and is computationally more efficient. We discuss the model's properties, including its limits and (ordinary and quantum-affine) symmetries. In particular we introduce the appropriate notions of translational invariance and momentum. We review the model's exact spectrum found by Uglov for finite spin-chain length, which parallels the isotropic case up to level splitting due to the anisotropy. We also extend the partially isotropic model to higher rank, with $SU(n)$ `spins', for which the spectrum is determined by $\mathfrak{sl}_n$-motifs.
\end{abstract}

\maketitle

\noindent The Haldane--Shastry model (HSM)~\cite{Hal_88,Sha_88} is a long-range spin chain with pairwise $1/r^2$-exchange interactions. A salient feature is its invariance under the Yangian of $\mathfrak{sl}_2$ for finite spin-chain length~\cite{HH+_92,BG+_93}, explaining in part~\cite{FG_15} the high degeneracies observed in the spectrum~\cite{Hal_88}, coming in representations of the Yangian. This infinite-dimensional symmetry algebra renders the HSM exactly solvable.

One naturally wonders whether the $\mathfrak{su}_2$-symmetry can be broken to get a \emph{partially} isotropic (\textsc{xxz}-like) version of the HSM with quantum-affine symmetry. Uglov~\cite{Ugl_95u} provided a positive answer, building on~\cite{BG+_93,TH_95}, yielding a rather complicated Hamiltonian. The work was never published and appears to have been forgotten.

With this text we wish to revive and continue Uglov's work. We present a novel expression for the Hamiltonian that parallels the structure of the HSM. We give a down-to-earth review of its properties, introduce the appropriate notion of translation invariance, and review the exact spectrum~\cite{Ugl_95u}. We prove that our formula equals that of Uglov. This shows that Uglov's Hamiltonian is hermitian in the `easy-axis' case where the anisotropy is Ising-like, corresponding to $|\Delta|\geq 1$ for the Heisenberg spin chain. We moreover generalize the model to the partially isotropic version of the multicomponent HSM with $n$~`colours'~\cite{Kaw_92,*HH_92,*HH_93}. Further details will be given elsewhere~\cite{Lam_18u_maths}.

\section{The model}

\noindent Consider a ring with $L$ equally spaced sites with spin~$1/2$. The Hilbert space $\mathcal{H} = (\mathbb{C}^2)^{\otimes L}$ is the tensor product of for each site one spin-1/2 space with basis $\ket{\uparrow}$ and $\ket{\downarrow}$. The Hamiltonian can be written in the HSM-like form
	\begin{equation} \label{eq:Ham_new}
	H = {-J} \, \sum_{i<j}^L V(i-j) \, S_{[i,j]} \, ,
	\end{equation}
where the sum ranges over all pairs of sites. The constituents of \eqref{eq:Ham_new} are as follows. $J$ is a coupling constant, with $J>0$ ($J<0$) favouring (anti)ferromagnetic order. The potential~$V$ is a `point-splitting' of the inverse-square pair potential of the HSM, where the chord distance~$r$ between the sites is deformed by a (real) anisotropy parameter~$\gamma$ that here acts as a regulator for the \textsc{uv} (short-distance) divergence of the HSM potential:
\begin{subequations}
	\label{eq:potential_new}
	\begin{gather}
	V(k) = \frac{1}{r_+(k)\,r_-(k)} \, , \quad r_\pm(k) = 2\, \sin\Bigl(\frac{\pi\,k}{L} \pm \I\,\gamma\Bigr) \, .
\intertext{For the HSM it is customary to view the sites as positioned at $z_k = \E^{2\pi\I\,k/L}$. If further $q=\E^\gamma$ we can write}
	V(i,j) = \frac{z_i \, z_j}{(q\,z_i-q^{-1}\,z_j)(q\,z_j-q^{-1}\,z_i)} \, .
	\end{gather}
\end{subequations}

The parameters~$z_k$ also enter the long-range interaction operators~$S_{[i,j]}$ in \eqref{eq:Ham_new}. These are built from the \textit{R}-matrix, which is defined on $\ket{\uparrow\uparrow}, \ket{\uparrow\downarrow}, \ket{\downarrow\uparrow}, \ket{\downarrow\downarrow}$ by
	\begin{equation} \label{eq:R_matrix}
	\check{R}(u) = \begin{pmatrix}
	1 & 0 & 0 & 0 \\
	0 & u\,g(u) & f(u) & 0 \\
	0 & f(u) & g(u) & 0 \\
	0 & 0 & 0 & 1
	\end{pmatrix} , \quad \begin{aligned}
	& f(u)=\frac{u-1}{q\,u-q^{-1}} \, , \\ 
	& g(u) = \frac{q-q^{-1}}{q\,u-q^{-1}} \, .
	\end{aligned}
	\end{equation}
$\check{R}(u) P$, with $P$ the permutation matrix, is the fundamental object for the treatment of the \textsc{xxz} model via the quantum inverse-scattering method (\textsc{qism}) \cite{[{See e.g.\ }] [{, and references therein.}]Lam_14}, here in multiplicative notation ($u=\E^{2\lambda}$, $q=\E^\gamma$) and the `homogeneous picture'~\cite{[{Compare \unexpanded{$\bra{\uparrow\downarrow} \check{R}(u) \ket{\uparrow\downarrow}$} with \unexpanded{$\bra{\downarrow\uparrow} \check{R}(u) \ket{\downarrow\uparrow}$}. This ensures nontrivial `braid limits' \unexpanded{$u\to0,\infty$}, yielding the \unexpanded{$U_{q}(\mathfrak{sl}_{2})$} \unexpanded{$R$}-matrix. See also \textsection5.4 of }] [{}]JM_95,*[{{}\textsection10 of }] [{}]Fad_95u}. We have \cite{[{This form was foreseen in (16) of }] [{, although the Hamiltonian and quantum-affine symmetry constructed there are different. }]HS_96,*[{These `braid translations' originate in }] [{}]MS_94}
\begin{subequations}
	\label{eq:S_ij}
	\begin{gather}
	\begin{aligned}
	S_{[i,j]} = \ & \biggl(\,\ordprod_{j>k>i} \!\! \check{R}_{k,k+1}(z_k/z_j)\biggr) \, (q-q^{-1})\,\check{R}_{i,i+1}'(1) \\
	& \times \biggl(\,\ordprodopp_{i<k<j} \!\! \check{R}_{k,k+1}(z_j/z_k)\biggr) \, , \qquad\quad i<j \, ,
	\end{aligned}
\intertext{where the products run over $k$ and the harpoons specify the ordering. This expression can be understood as}
	S_{[i,j]} =
	\tikz[baseline={([yshift=-.5*10pt*0.6]current bounding box.center)},	xscale=0.5,yscale=0.25,font=\footnotesize]{
		\draw[rounded corners=3pt] (8,0) node[below]{$z_j$} -- (8,.5) -- (5,3.5) -- (5,4);
		\draw[rounded corners=3pt] (7,0) node[below]{$z_{j-1}$} -- (7,.5) -- (8,1.5) -- (8,6.5) -- (7,7.5) -- (7,8) node[above]{$z_{j-1}$};
		\draw[rounded corners=3pt] (6,0) node[below]{$\cdots$} -- (6,1.5) -- (7,2.5) -- (7,5.5) -- (6,6.5) -- (6,8) node[above]{$\cdots{\vphantom{z_j}}$};
		\draw[rounded corners=3pt] (5,0) node[below]{$z_{i+1}$} -- (5,2.5) -- (6,3.5) -- (6,4.5) -- (5,5.5) -- (5,8) node[above]{$z_{i+1}\vphantom{z_j}$};
		\draw[rounded corners=3pt] (5,4) -- (5,4.5) -- (8,7.5) -- (8,8) node[above]{$z_j$};
		\draw[rounded corners=3pt] (4,0) node[below]{$z_i$} -- (4,8) node[above]{$z_i\vphantom{z_j}$};
		\draw[style={decorate, decoration={zigzag,amplitude=.5mm,segment length=2mm}}] (4,4) -- (5,4);
		\foreach \x in {-1,...,1} \draw (3.5+.2*\x,4) node{$\cdot\mathstrut$};
		\foreach \x in {4,...,8} \draw[->] (\x,8) -- (\x,8.05);	
		\foreach \x in {-1,...,1} \draw (8.5+.2*\x,4) node{$\cdot\mathstrut$};	
	} 
	, \quad \begin{array}{l}
	\tikz[baseline={([yshift=-.5*10pt*0.6]current bounding box.center)}, 	xscale=0.5,yscale=0.25,font=\footnotesize]{
		\draw[rounded corners=3pt,->] (1,0) node[below]{$v$} -- (1,.5) -- (0,1.5) -- (0,2) node[above]{$v$};
		\draw[rounded corners=3pt,->] (0,0) node[below]{$u$} -- (0,.5) -- (1,1.5) -- (1,2) node[above]{$u$};
	} \!\!= \check{R}(v/u) \, , \\ \\
	\tikz[baseline={([yshift=-.5*10pt*0.6]current bounding box.center)}, 	xscale=0.5,yscale=0.25,font=\footnotesize]{
		\draw[->] (1,0) node[below]{$u$} -- (1,2) node[above]{$u$};
		\draw[->] (2,0) node[below]{$v$} -- (2,2) node[above]{$v$};
		\draw[style={decorate, decoration={zigzag,amplitude=.5mm,segment length=2mm}}] (1,1) -- (2,1);
	} \!\!= (q\,{-}\,q^{-1}) \, \mathrlap{ \check{R}'(1) \, . }
	\end{array} \label{eq:Sij_graphical}
	\end{gather}
\end{subequations}
The notation `$[i,j]$' as an interval will make sense soon.

The physical picture is as follows: the spin at site~$j$ is transported down to~$i\,{+}\,1$ to interact with the spin at site~$i$ and then brought back to~$j$. The transport uses the \textit{R}-matrix~\eqref{eq:R_matrix}, where the spins take along their parameter~$z_k$ as in \eqref{eq:Sij_graphical}. The nearest-neighbour interaction 
	\begin{equation} \label{eq:R'}
	(q\,{-}\,q^{-1})\,\check{R}'(1) = \begin{pmatrix}
	0 & 0 & 0 & 0 \\
	0 & {-q}^{-1} & 1 & 0 \\
	0 & 1 & {-q} & 0 \\
	0 & 0 & 0 & 0
	\end{pmatrix}
	\end{equation}
equals $-(q\,{+}\,q^{-1})$ times the \textit{q}-antisymmetrizer (projector onto the \textit{q}-singlet). The appearance of $\check{R}'(1)$ is familiar from the \textsc{qism}~\cite{[{See e.g.\ }] [{, and references therein.}]Lam_14}: besides a factor of two, \eqref{eq:R'} only differs from the usual \textsc{xxz} interaction $\sigma^x_i \sigma^x_i + \sigma^y_i \sigma^y_{i+1} + \Delta \, (\sigma^z_i \sigma^z_{i+1} -\id)$, $\Delta=(q+q^{-1})/2$, because of the `homogeneous picture'. Thus, \eqref{eq:S_ij} should be compared with the decomposition
	\begin{equation}
	P_{ij} -\id = P_{j-1,j} \cdots P_{i+1,i+2} (P_{i,i+1}-\id) P_{i+1,i+2} \cdots P_{j-1,j}
	\end{equation}
for the long-range interactions of the HSM.

\subsection{Key properties}

\noindent To show that \eqref{eq:Ham_new} is indeed the appropriate generalization of the HSM we list its most important properties.

\paragraph{Isotropic limit.} As $q \,{=}\, \E^\gamma \,{\to}\, 1$ we obtain the HSM. Indeed, \eqref{eq:potential_new} clearly has the right limit, while $\check{R}(u) \,{\to}\, P$ and $(q\,{-}\,q^{-1})\,\check{R}'(1) \,{\to}\, P\,{-}\,\id$ so $S_{[i,j]} \to P_{ij} \,{-}\, \id$.

\paragraph{Partial isotropy.} Spin-$z$ is conserved: \eqref{eq:Ham_new} commutes with $S^z = \sum \id^{\!\otimes(k-1)} \otimes \, (\sigma^z\!/2) \otimes \id^{\otimes(L-k)}$. This symmetry is inherited through \eqref{eq:S_ij} from the \textit{R}-matrix.

\paragraph{Quantum-affine symmetry.} Crucially, for each $L$ \eqref{eq:Ham_new} commutes with the action of an infinite-dimensional quantum group: the partial anisotropy deforms the Yangian symmetry of the HSM~\cite{HH+_92,BG+_93} to quantum-affine~$\mathfrak{sl}_2$, which is usually denoted by \Uqaff. Invariance under \Uqaff\ is guaranteed by Uglov's derivation of the Hamiltonian following~\cite{BG+_93,TH_95} by `freezing' a dynamical spin model, as will be reviewed elsewhere~\cite{Lam_18u_maths}. This should be contrasted with the Heisenberg models, which only enjoy such symmetries as $L\to\infty$~\cite{Ber_92,*DF+_93}. 

At `level zero' (degree zero) the quantum-affine symmetries contain quantum~$\mathfrak{sl}_2$, denoted by $U_q(\mathfrak{sl}_2)$, consisting of $S^z$ together with the \textit{q}-ladder operators
\begin{subequations} \label{eq:q_sl2}
	\begin{gather}
	S^\pm_q = \sum_{k=1}^L (q^{\sigma^z\!/2})^{\otimes(k-1)} \otimes \sigma^\pm \otimes (q^{-\sigma^z\!/2})^{\otimes(L-k)} \, ,
\intertext{where $\sigma^\pm = (\sigma^x \,{\pm}\, \I\, \sigma^y)/2$, with commutation relations}
	[S^z,S^\pm_q] = \pm S^\pm_q \, , \quad [S^+_q,S^-_q] = \frac{q^{2\,S^z}{-}\,q^{-2\,S^z}}{q\,{-}\,q^{-1}} \, .
	\end{gather}
\end{subequations}
Thus, despite the \emph{partial} isotropy, multiplets have a descendant structure as in isotropic models. The symmetries~\eqref{eq:q_sl2} also appear for certain open and quasiperiodic Heisenberg spin chains~\cite{PS_90,*KS_91}. The present model, however, has \emph{many} more symmetries. We will briefly get back to the `higher' symmetries contained in \Uqaff\ at the end of this section. Instead we will focus on the practical consequence of this infinite-dimensional symmetry algebra: the highly degenerate spectrum.

\paragraph{Additive energies.} Besides its high degeneracies, the spectrum of the HSM is special in that it is very regular: its energies are quantized as half-integer multiples of~$J$, though not all such multiples occur~\cite{Hal_88}. This regular pattern is the consequence of additivity of the energy together with a simple dispersion relation. We will see that $q\neq\pm1$ deforms the half-integrality of the energy, yet the spectrum remains strictly additive.

\subsection{Further properties} 

\noindent The preceding properties say that we are dealing with the correct generalization of the HSM. Before turning to the spectrum we discuss a few more properties and quirks of the partially isotropic HSM.

\paragraph{Hermiticity.} The Hamiltonian is hermitian for $q$ real. This is clear from~\eqref{eq:Ham_new}--\eqref{eq:S_ij}, yet not at all obvious from Uglov's expression; see below. In terms of the \textsc{xxz} model's parameter, real~$q$ corresponds to the massive (easy-axis) regime $|\Delta|>1$. Replacing $q$ by $-q$ yields an overall minus sign for the energy; for even $L$ it amounts to conjugating $H$ by either of $(\sigma^z\!\otimes\id)^{\otimes L/2}$ and $(\id\otimes\,\sigma^z)^{\otimes L/2}$, again up to an overall sign.

\paragraph{Multi-spin interactions.} Despite the pairwise form of \eqref{eq:Ham_new} the partially anisotropic long-range interactions~\eqref{eq:S_ij} affect all intermediate spins when $q\neq \pm 1$, taking into account interactions between multiple spins. This structure is more manifest in Uglov's formula. It is also the reason why we write the subscript of \eqref{eq:S_ij} as an interval (except when $j=i+1$). Physically such long-ranged multi-spin interactions are acceptable, and do indeed occur in any real material, as long as their strength falls off sufficiently rapidly. We intend to investigate this issue of locality in the near future.

\paragraph{Parity violating.} For $q\neq\pm1$ \eqref{eq:Ham_new} is \emph{not} invariant under parity reversal ($i\mapsto L-i+1$), as is already suggested by the asymmetric roles played by $i$ and $j$ in \eqref{eq:S_ij}. There is, however, a `\textsc{cpt}-invariance' under simultaneously reversing $\ket{\uparrow}\leftrightarrow\ket{\downarrow}$, $z_i \mapsto z_i^{-1} \,{=}\, z_{L-i+1}$
and $q\mapsto q^{-1}$.

\paragraph{\textit{q}-homogeneity.} The potential~\eqref{eq:potential_new} is translationally invariant, yet for $q\neq{\pm 1}$ the operators~\eqref{eq:S_ij} are \emph{not}. This, and the absence of periodicity in the usual sense, is particularly clear comparing $S_{[1,L]}$ with any $S_{i,i+1}$. However, there is a \textit{q}-analogue of homogeneity: $H$ commutes~\cite{Lam_18u_maths} with the (unitary) \textit{q}-translation operator
\begin{equation} \label{eq:q-shift}
	U = \! \ordprod_{L>k\geq 1}\!\!\! \check{R}_{k,k+1}(z_{k+1}/z_1) \, = \!
	\tikz[baseline={([yshift=-.5*10pt*0.6]current bounding box.center)},
	xscale=0.5,yscale=0.25,font=\footnotesize]{
		\foreach \x in {1,...,4} \draw[rounded corners=3pt] (\x,0) -- ++ (0,\x-.5) -- ++ (-1,1) -- (\x-1,5);
		\node at (0,5) [above]{$z_2$};
		\node at (1,0) [below]{$z_2$};
		\node at (2.5,0) [below]{$\cdots$};
		\node at (1.5,5) [above]{$\cdots$};
		\node at (3,5) [above]{$z_L$};
		\node at (4,0) [below]{$z_L$};
		\draw[rounded corners=3pt] (0,0) node[below]{$z_1$} -- (0,.5) -- (4,4.5) -- (4,5) node[above]{$z_1$};
		\foreach \x in {0,...,4} \draw[->] (\x,5) -- (\x,5.05);	
	} ,
\end{equation}
which reduces to the usual (left) shift operator as $q \,{\to}\, 1$. Thus there is a \textit{q}-analogue of (crystal) momentum, determined by the eigenvalue $\E^{\I\,p}$ of~$U$ and quantized as $p = \frac{2\pi}{L} \, m\!\!\mod 2\pi$ by the \textit{q}-periodic boundary conditions $U^L = \id$, which holds by the Yang--Baxter equation for \eqref{eq:R_matrix}. Similar modified translation operators appear for inhomogeneous quantum-integrable spin chains~\cite{[{See e.g.\ }] [{, and references therein.}]Lam_14} and $2d$ electrons moving in a transverse magnetic field.

\subsection{Exact spectrum}

\begin{figure}[t] \centering
	\includegraphics[width=1.0\linewidth,clip=]{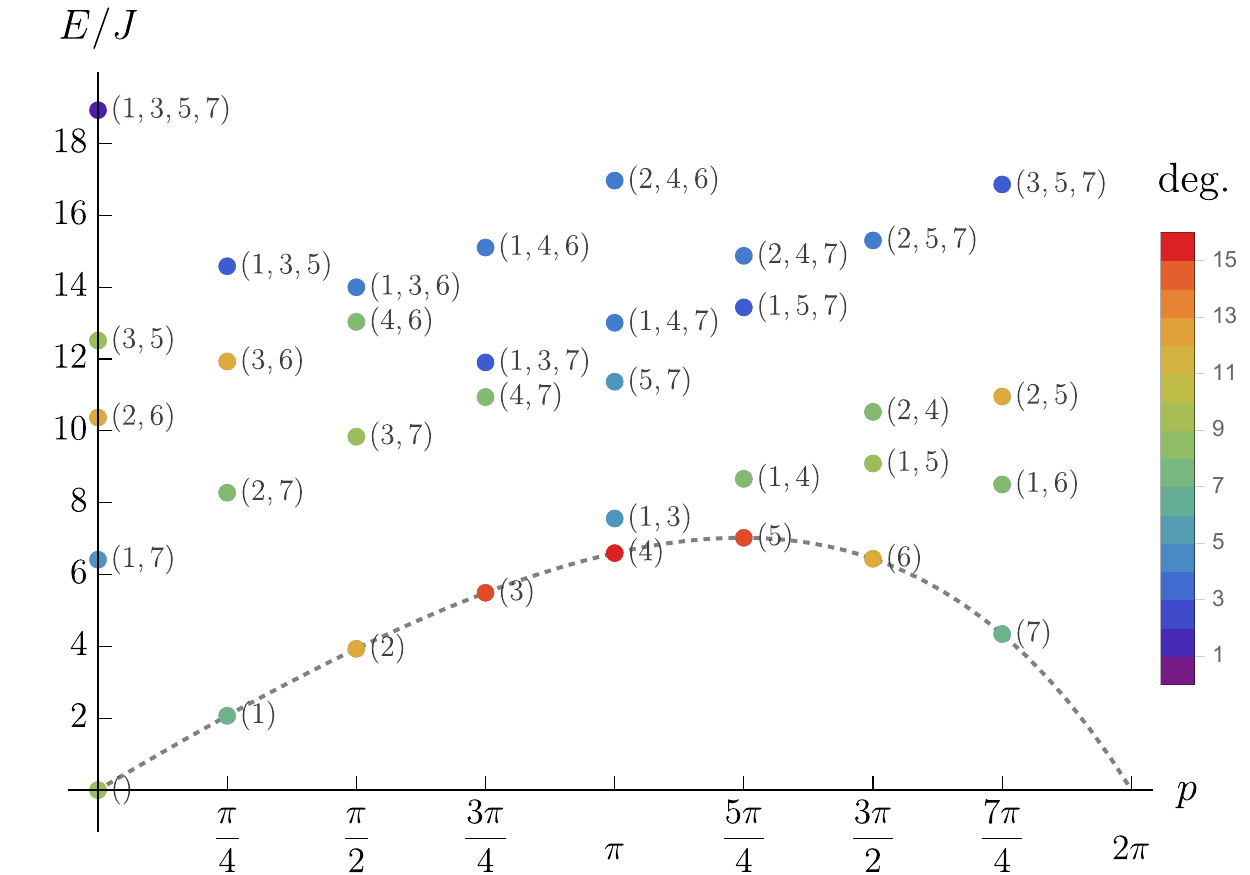}
	\caption{The full spectrum for $\mathfrak{su}_2$, $L=8$ and $\gamma=1/5$. Each dot represents a multiplet, labelled by its motif and with colour indicating its degeneracy. The dotted curve is the (off-shell) magnon dispersion~$\varepsilon(p)/J$. The additivity of the \textit{q}-momentum (mod~$2\pi$) and energy is manifest.}
	\label{fig:E_p_plot}
\end{figure}

\noindent Due to the quantum-affine symmetries the exact spectrum is known explicitly for any spin-chain length~$L$. Let us review the results of \cite{Ugl_95u} from a more physical viewpoint. We have verified the following numerically with random values for~$q=\E^\gamma$ for $L\leq 16$. While going through the following the reader may wish to consult the examples in Table~\ref{tb:L=4} and Fig.~\ref{fig:E_p_plot}.

Just as for the HSM the combinatorial structure of the spectrum is given by `motifs'~\cite{HH+_92}. For a given length~$L$ a \emph{motif} is a sequence $(m_r)_r$ of increasing integers $1\leq m_r\leq L\,{-}\,1$ differing by more than one: $m_{r+1} > m_r\,{+}\,1$. A motif can be represented by $\tfrac12\,{+}\,(L\,{-}\,1)\,{+}\,\tfrac12$ (semi)circles, 
$\tikz[baseline={([yshift=-.5*10pt*0.6]current bounding box.center)},
scale=0.2,font=\footnotesize]{
	\draw (0,0) arc (-90:90:.5); 
	\draw (1.25,.5) circle (.5); 
	\draw (2.5,.5) circle (.5); 
	\foreach \x in {-1,...,1} \draw (3.75+.4*\x,.5) node{$\cdot\mathstrut$};
	\draw (5,.5) circle (.5); 
	\draw (6.25,0) arc (270:90:.5);
}$\,, 
where for every $r$ the $m_r$th full circle is filled; then there are no adjacent $\tikz[baseline={([yshift=-.5*10pt*0.6]current bounding box.center)},
scale=0.2,font=\footnotesize]{
	\draw[fill=black] (0,.5) circle (.5);
}$\,s. This pattern may be interpreted as a `generalized Pauli principle'~\cite{Hal_91b}.

Each motif corresponds to a multiplet that has linear \textit{q}-momentum $p = \sum_r 2\pi\,m_r/L \!\!\mod 2\pi$, and additive energy $E = \sum_r \varepsilon(m_r)$ with dispersion relation
\begin{align} \label{eq:dispersion}
	\varepsilon(m) & = J \, \frac{1}{q-q^{-1}} \, \Bigl(m - L \, \frac{q^m \, [m]_q}{q^L \, [L]_q} \Bigr) \\ 
	& = \frac{J}{[L]_q} \, \sum_{n=1}^{L-1} \min\bigl(n\,m,(L-m)(L-n)\bigr)\, q^{L-2n} \, , \nonumber
\end{align}
with \textit{q}-integers defined as $[L]_q = (q^L{-}\,q^{-L})/(q\,{-}\,q^{-1})$. The anisotropy tilts the dispersion, cf.~Fig.~\ref{fig:E_p_plot}, causing a level-splitting of the HSM's spectrum. The quadratic relation $\varepsilon_\text{HS}(m) = J \, m\,(L\,{-}\,m)/2$ is recovered as $q\to 1$. 
Mirror-image motifs yield equal energy only for $q=\pm1$; generically motifs and energies are in one-to-one correspondence: there are no `accidental' degeneracies, cf.~\cite{FG_15}.

Recall that the quantum-integrable Heisenberg model has `functionally additive' energies in terms of the quasimomenta. Its spectrum, however, is rather more complicated because the Bethe-ansatz equations determining the quasimomenta admit complex solutions, which may (asymptotically) be interpreted as bound states of magnons. Instead, motifs give a simple combinatorial rule for the allowed quasimomenta~$p_r = 2\pi\, m_r/L$ and occurring energies, which together with the additivity means that the spectrum describes an ideal gas of quasiparticles interacting through their statistics only.

Motifs also characterize the degeneracy, and in fact the \Uqaff-content of the multiplet. Namely, the multiplet is the tensor product of the following factors: first replace each 
$\tikz[baseline={([yshift=-.5*10pt*0.6]current bounding box.center)},
scale=0.2,font=\footnotesize]{
	\draw (0,0) arc (-90:90:.5); 
	\draw[fill=black] (1.25,.5) circle (.5); 
	\draw (2.5,0) arc (270:90:.5);
}$ 
by a \textit{q}-singlet 
$\tikz[baseline={([yshift=-.5*10pt*0.6]current bounding box.center)},
scale=0.2,font=\footnotesize]{
	\draw (0,0) rectangle (1,1) rectangle (0,2);
}$\,, 
and then every remaining string of $\tfrac12\,{+}\,(n\,{-}\,1)\,{+}\,\tfrac12$ open (semi)circles
$\tikz[baseline={([yshift=-.5*10pt*0.6]current bounding box.center)},
scale=0.2,font=\footnotesize]{
	\draw (0,0) arc (-90:90:.5); 
	\draw (1.25,.5) circle (.5); 
	\foreach \x in {-1,...,1} \draw (2.5+.4*\x,.5) node{$\cdot\mathstrut$};
	\draw (3.75,.5) circle (.5); 
	\draw (5,0) arc (270:90:.5);
}$
by a \textit{q}-symmetric irrep
$\tikz[baseline={([yshift=-.5*10pt*0.6]current bounding box.center)},
scale=0.2,font=\footnotesize]{
	\draw (0,0) rectangle (1,1) rectangle (2,0);
	\draw (2.5,0) rectangle (3.5,1) rectangle (4.5,0);
	\draw[white,fill=white] (1.5,-.05) rectangle (3,1.05);
	\foreach \x in {-1,...,1} \draw (2.25+.4*\x,.5) node{$\cdot\mathstrut$};
}$ with $n$ boxes. This tensor product may be \emph{re}ducible as a $U_q(\mathfrak{sl}_2)$-representation, but not for \Uqaff: here the `higher' symmetries show up in the spectrum. The degeneracies can be counted as for $\mathfrak{su}_2$. The spectrum is less degenerate than for the HSM, yet much more than for the Heisenberg models.

\begin{table}
	\begin{tabular}{cccccccc} \toprule
		\multicolumn{2}{c}{motif} & q.-a.~irrep & \textit{q}-spin & deg. & $p$ & & $E\times [4]_q/J$ \\ \midrule
		$(\,)$ & $\tikz[baseline={([yshift=-.5*10pt*0.6]current bounding box.center)}, scale=0.2,font=\footnotesize]{
			\draw (0,0) arc (-90:90:.5); 
			\draw (1.25,.5) circle (.5); 
			\draw (2.5,.5) circle (.5); 
			\draw (3.75,.5) circle (.5); 
			\draw (5,0) arc (270:90:.5);
		}$ & $\tikz[baseline={([yshift=-.5*10pt*0.6]current bounding box.center)}, scale=0.175,font=\footnotesize]{
			\draw (0,0) rectangle (1,1) rectangle (2,0) rectangle (3,1) rectangle (4,0);
		}$ & $\mathbf{2}$ & $5$ & $0$ & & $0$ \\
		$(1)$ & $\tikz[baseline={([yshift=-.5*10pt*0.6]current bounding box.center)}, scale=0.2,font=\footnotesize]{
			\draw (0,0) arc (-90:90:.5); 
			\draw[fill=black] (1.25,.5) circle (.5);
			\draw (2.5,.5) circle (.5); 
			\draw (3.75,.5) circle (.5); 
			\draw (5,0) arc (270:90:.5);
		}$ & $\tikz[baseline={([yshift=-.5*10pt*0.6]current bounding box.center)}, scale=0.175,font=\footnotesize]{
			\draw (0,0) rectangle (1,1) rectangle (0,2);
		} \otimes \tikz[baseline={([yshift=-.5*10pt*0.6]current bounding box.center)}, scale=0.175,font=\footnotesize]{
			\draw (0,0) rectangle (1,1) rectangle (2,0);
		}$ & $\mathbf{1}$ & $3$ & $2\pi/4$ & & $\hphantom{2\,} q^2\,{+}\,2\,{+}\,3\,q^{-2}$ \\
		$(2)$ & $\tikz[baseline={([yshift=-.5*10pt*0.6]current bounding box.center)}, scale=0.2,font=\footnotesize]{
			\draw (0,0) arc (-90:90:.5); 
			\draw (1.25,.5) circle (.5); 
			\draw[fill=black] (2.5,.5) circle (.5); 
			\draw (3.75,.5) circle (.5); 
			\draw (5,0) arc (270:90:.5);
		}$ & $\tikz[baseline={([yshift=-.5*10pt*0.6]current bounding box.center)}, scale=0.175,font=\footnotesize]{
			\draw (0,0) rectangle (1,1);
		} \otimes \tikz[baseline={([yshift=-.5*10pt*0.6]current bounding box.center)}, scale=0.175,font=\footnotesize]{
			\draw (0,0) rectangle (1,1) rectangle (0,2);
		} \otimes \tikz[baseline={([yshift=-.5*10pt*0.6]current bounding box.center)}, scale=0.175,font=\footnotesize]{
			\draw (0,0) rectangle (1,1);
		} $ & $\mathbf{0}\,{\oplus}\,\mathbf{1}$& $4$ & $4\pi/4$ & & $2\,q^2\,{+}\,4\,{+}\,2\,q^{-2}$ \\
		$(3)$ & $\tikz[baseline={([yshift=-.5*10pt*0.6]current bounding box.center)}, scale=0.2,font=\footnotesize]{
			\draw (0,0) arc (-90:90:.5); 
			\draw (1.25,.5) circle (.5); 
			\draw (2.5,.5) circle (.5); 
			\draw[fill=black] (3.75,.5) circle (.5); 
			\draw (5,0) arc (270:90:.5);
		}$ & $\tikz[baseline={([yshift=-.5*10pt*0.6]current bounding box.center)}, scale=0.175,font=\footnotesize]{
			\draw (0,0) rectangle (1,1) rectangle (2,0);
		} \otimes \tikz[baseline={([yshift=-.5*10pt*0.6]current bounding box.center)}, scale=0.175,font=\footnotesize]{
			\draw (0,0) rectangle (1,1) rectangle (0,2);
		}$ & $\mathbf{1}$ & $3$ & $6\pi/4$ & & $3\,q^2\,{+}\,2\,{+}\,\hphantom{2\,}q^{-2}$ \\
		$(1,3)$ & $\tikz[baseline={([yshift=-.5*10pt*0.6]current bounding box.center)}, scale=0.2,font=\footnotesize]{
			\draw (0,0) arc (-90:90:.5); 
			\draw[fill=black] (1.25,.5) circle (.5); 
			\draw (2.5,.5) circle (.5); 
			\draw[fill=black] (3.75,.5) circle (.5); 
			\draw (5,0) arc (270:90:.5);
		}$ & $\tikz[baseline={([yshift=-.5*10pt*0.6]current bounding box.center)}, scale=0.175,font=\footnotesize]{
			\draw (0,0) rectangle (1,1) rectangle (0,2);
		} \otimes \tikz[baseline={([yshift=-.5*10pt*0.6]current bounding box.center)}, scale=0.175,font=\footnotesize]{
			\draw (0,0) rectangle (1,1) rectangle (0,2);
		}$ & $\mathbf{0}$ & $1$ & $0$ & & $4\,q^2\,{+}\,4\,{+}\,4\,q^{-2}$ \\
		\bottomrule
		\hline
	\end{tabular}
	\caption{The exact spectrum for $\mathfrak{su}_2$ at $L=4$: motifs, the quantum-affine and \textit{q}-spin content of the corresponding multiplet, its degeneracy, \textit{q}-momentum ($\!\!\!\!\mod2\pi$) and energy. (The coinciding coefficients for the \textsc{af} energy are accidental.)}
	\label{tb:L=4}
\end{table}

The empty motif~$(\,)$ corresponds to the \textit{q}-ferromagnetic multiplet, consisting of $\ket{\uparrow{\cdots}\uparrow}$ and its `level-zero' \textit{q}-descendants $(S^-_q)^M \, \ket{\uparrow{\cdots}\uparrow}$ for $1\leq M\leq L$, with $S^-_q$ from \eqref{eq:q_sl2}. The $M$th descendant has $S^z=L/2-M$. Next, each motif of the form $(m)$ describes a \textit{q}-magnon with $p\neq0$ and $S^z=L/2-1$ together with its \textit{q}-descendants. $S^-_q$ produces $L-2$ `level-zero' descendants, and there are additional `higher-level' (`affine') descendants for $2\leq m\leq L-2$. This descendant structure is compatible with the value $\varepsilon(0)=0$; cf.~$\varepsilon_\textsc{xxz}(0) = J(\Delta-1)$ vanishing only at the isotropic point $\Delta=1$.
 
At the other end of the spectrum we find for even~$L$ the singlet $\tikz[baseline={([yshift=-.5*10pt*0.6]current bounding box.center)}, scale=0.175,font=\footnotesize]{
	\draw (0,0) rectangle (1,1) rectangle (0,2);
} \otimes \cdots\otimes \tikz[baseline={([yshift=-.5*10pt*0.6]current bounding box.center)}, scale=0.175,font=\footnotesize]{
	\draw (0,0) rectangle (1,1) rectangle (0,2);
}$ corresponding to the antiferromagnetic (\textsc{af}) motif $(1,3,\To,L-1)$ with $p=L\pi/2\!\!\mod2\pi$ and $S^z=0$. When $J<0$ this is the unique ground state. For odd $L$ the \textsc{af} motif is not allowed and there is at least one $\tikz[baseline={([yshift=-.5*10pt*0.6]current bounding box.center)},
scale=0.2,font=\footnotesize]{
\draw (0,0) arc (-90:90:.5);
\draw (1.25,0) arc (270:90:.5);
}$, yielding a doublet that we interpret as a \textit{q}-spinon ($S^z=1/2$) and its `level-zero' \textit{q}-descendant. Back at even $L$ the excitations over the \textit{q}-\textsc{af} vacuum are described by motifs with two $\tikz[baseline={([yshift=-.5*10pt*0.6]current bounding box.center)}, scale=0.2,font=\footnotesize]{
\draw (0,0) arc (-90:90:.5);
\draw (1.25,0) arc (270:90:.5);
}$\,s that may or may not be adjacent and describe two \textit{q}-spinons. The $L$ motifs yielding for each \textit{q}-momentum the lowest excitation are obtained from the \textsc{af}~motif by removing $m=1$ or $m=L-1$ and then step by step moving the remaining $\tikz[baseline={([yshift=-.5*10pt*0.6]current bounding box.center)}, scale=0.2,font=\footnotesize]{
	\draw[fill=black] (0,.5) circle (.5);
}$\,s towards the new $\tikz[baseline={([yshift=-.5*10pt*0.6]current bounding box.center)}, scale=0.2,font=\footnotesize]{
\draw (0,.5) circle (.5);
}$, cf.~Fig.~\ref{fig:E_p_plot}.

\subsection{Quantum-affine symmetries}

\noindent Let us sketch the (level $c=0$) action of quantum-affine~$\mathfrak{sl}_2$, which goes via a monodromy matrix as usual~\cite{[{See e.g.\ }] [{, and references therein.}]Lam_14}, but is rather more involved than the Yangian symmetry of the HSM. For a multiplet, characterized by some motif, consider
	\begin{equation} \label{eq:monodromy_model}
	T_a(u) = \!\ordprodopp_{1\leq k\leq L}\!\!\! L_{ak}(q^{\mu_k}u) \, , \quad L(u)= \frac{1}{f(u)} \, \check{R}(u) \, P \, ,
	\end{equation}
where $\mu_k = 2\,k \,{-}\,L\,{-}\,1$ unless $k$ or $k\,{-}\,1$ is contained in the motif, in which case the values are swapped: $\mu_{m_r} = 2 \, m_r\,{-}\,L\,{+}\,1$, $\mu_{m_r+1} = 2\, m_r\,{-}\,L\,{-}\,1$ for all $r$. The operator \eqref{eq:monodromy_model} obeys the \textit{RTT}-relations with \textit{R}-matrix $\bar{R}(u) = \check{R}(u)\, P$. Only for the empty motif, however, do the four `quantum operators' contained in \eqref{eq:monodromy_model} commute with $H$ on the corresponding eigenspace. The actual monodromy matrix is modelled on \eqref{eq:monodromy_model} and involves the values of a special case of nonsymmetric Macdonald polynomials at the point $\{z_k = \E^{2\pi\I\, k/L} \}$~\cite{Ugl_95u}. Expansions in $u^{\pm1}$ yield infinite towers of `higher' symmetries, with \eqref{eq:q_sl2} at zeroth order. In fact \eqref{eq:q_sl2} can already be found from \eqref{eq:monodromy_model}, which becomes independent of the motif in the `braid limits' $u\to0$ and $u\to\infty$.

\section{Uglov's formula} 

\noindent Let us briefly recall the Hamiltonian found by Uglov~\cite{Ugl_95u}. Up to a rescaling it can be written as
\begin{equation} \label{eq:Ham_Uglov}
\tilde{H} = {-J} \sum_{N=2}^L \, \sum_{i_1 < \cdots < i_N} \!\!\!\! \tilde{V}(i_1,\To,i_N) \, (Y_{[i_1,\To,i_N]} - \id) \, .
\end{equation}
Here an $N$-point interaction has potential
\begin{align} \label{eq:potential_Uglov}
\tilde{V}(i_1,\To,i_N) = \ & \frac{L/[L]_q}{(q-q^{-1})^2} \\
& \times \prod_{n = 1}^N \frac{g(z_{i_n}/z_{i_{n+1}})}{f(z_{i_n}/z_{i_{n+1}})} \! \prod_{k=i_n+1}^{i_{n+1}-1} \! \frac{1}{f(z_k/z_{i_n})} \, , \nonumber
\end{align}
with $[L]_q$ given below~\eqref{eq:dispersion}, and $i_{N+1} \equiv i_1$ and $z_{L+1} \equiv z_1$. The operator $Y_{[i_1,\To,i_N]}$ is given by
\begin{subequations} \label{eq:Y_ij}
	\begin{gather}
	\begin{aligned}
	Y_{[i_1,\To,i_N]} = \ & \ordprodopp_{1\leq n<N} \!\! Y_{[i_n,i_{n+1}]}(z_{i_n},z_{i_N}) \, , \\
	Y_{[i,j]}(u,v) = \ & \biggl(\,\ordprod_{j>k>i} \!\! \check{R}_{k,k+1}(z_k/u)\biggr) \, \check{R}_{i,i+1}(v/u) \\
	& \times \biggl(\,\ordprodopp_{i<k<j} \!\! \check{R}_{k,k+1}(v/z_k)\biggr) \, , \qquad i<j \, .
	\end{aligned}
	\intertext{In the diagrammatic notation from \eqref{eq:Sij_graphical} a typical example assumes the form}
	Y_{[i_1,i_2,i_3,i_4]} =
	\tikz[baseline={([yshift=-.5*10pt*0.6]current bounding box.center)},
	xscale=0.5,yscale=0.3,font=\footnotesize]{
		\draw[rounded corners=4pt] (8,0) node[below]{$z_{i_4}$} -- (8,.5) -- (2,6.5) -- (2,9) node[above]{$z_{i_4}$};
		\draw[rounded corners=4pt]
		(1,0) -- (1,9)
		(3,0) -- (3,4.5) -- (4,5.5) -- (4,6.5) -- (3,7.5) -- (3,9)
		(4,0) -- (4,3.5) -- (5,4.5) -- (5,7.5) -- (4,8.5) -- (4,9)
		(7,0) -- (7,.5) -- (8,1.5) -- (8,2.5) -- (7,3.5) -- (7,9)
		(9,0) -- (9,9)
		;
		\draw[rounded corners=4pt]
		(2,0) node[below]{$z_{i_1}$} -- (2,5.5) -- (5,8.5) -- (5,9) node[above]{$z_{i_1}$}
		(5,0) node[below]{$z_{i_2}$} -- (5,2.5) -- (6,3.5) -- (6,9) node[above]{$z_{i_2}$}
		(6,0) node[below]{$z_{i_3}$} -- (6,1.5) -- (8,3.5) -- (8,9) node[above]{$z_{i_3}$}
		;
		\foreach \x in {-1,...,1} \draw (.25+.2*\x,4.5) node{$\cdot\mathstrut$};
		\foreach \x in {1,...,9} \draw[->] (\x,9) -- (\x,9.05);	
		\foreach \x in {-1,...,1} \draw (9.75+.2*\x,4.5) node{$\cdot\mathstrut$};	
	} \, .
	\end{gather}
\end{subequations}
Note that $Y_{[1,\To,L]} = U^{-1}$ is the inverse of the \textit{q}-trans\-lation operator. As $q\to1$ \eqref{eq:Y_ij} just becomes the cyclic permutation of the spins at $i_1,i_2,\To,i_N$, though only the terms with $N=2$ actually survive this limit due to~\eqref{eq:potential_Uglov}.

Our formula is computationally much more efficient; on a laptop we have numerically obtained the full spectrum of \eqref{eq:Ham_new} for $L\leq 16$, as opposed to $L\leq 12$ for \eqref{eq:Ham_Uglov}.

\subsection{Sketch of proof of equality}

\noindent The `minimal hermitian constituents' of \eqref{eq:Ham_Uglov} consist of all terms with fixed $i_1 = i$ and $i_N = j$:
\begin{equation} \label{eq:h_ij}
	h_{[i,j]} = {-J} \! \sum_{N=2}^{j-i+1} \sum_{\substack{i_1<\cdots<i_N \\ i_1 =i,\, i_N=j}} \!\!\!\!\! \tilde{V}(i_1,\To,i_N) \, (Y_{[i_1,\To,i_N]}-\id) \, .
\end{equation}
In fact, we will show that this equals a single term of \eqref{eq:Ham_new}. 

The case $i=j\,{-}\,1$ straightforwardly follows from
\begin{equation}
	f(u)^{-1} \, \bigl(\check{R}(u) - \id\bigr) = (q-q^{-1}) \, \check{R}'(1) \label{eq:R-1toR'}
\end{equation}
along with $\prod_{k(\neq j)}^L f(z_k/z_j)^{-1} = [L]_q/L$, valid for any~$j$ since $z_k = \E^{2\pi\I k/L}$.

For $i<j-1$ we proceed recursively, at each step halving the number of terms as follows. Group the terms in \eqref{eq:h_ij} into pairs differing only in whether or not $j-1$ `partakes in the interaction':
\begin{align} \label{eq:h_ij_pairwise}
	h_{[i,j]} & = {-J} \!\! \sum_{n=0}^{j-i-1} \!\!\! \sum_{\substack{i_0<\cdots<i_n \\ i_0 =i,\, i_n<j-1}} \!\!\!\!\!\! \Bigl[ \tilde{V}(i_0,\To,i_n,j) \, (Y_{[i_0,\To,i_n,j]}-\id) \nonumber \\
	& + \tilde{V}(i_0,\To,i_n,j-1,j) \, (Y_{[i_0,\To,i_n,j-1,j]}-\id) \Bigr] \, ,
\end{align}
where $n$ is the number of interacting spins between $i$ and $j-1$. (Note the slight abuse of language: the `non-interacting' spins may still be affected by the transport.) We will combine the two terms in the summand of \eqref{eq:h_ij_pairwise} using \eqref{eq:R-1toR'} and
\begin{equation} \label{eq:Vtilde_cancel}
	\begin{aligned}
	& \biggl(1-\frac{f(z_{j-1}/z_{i_{n-1}})}{f(z_{j-1}/z_j)}\biggr) \, \tilde{V}(i_0,\To,i_n,j) \\
	& \qquad\qquad\qquad\qquad\ + \tilde{V}(i_0,\To,i_n,j-1,j) = 0 \, . 
	\end{aligned}
\end{equation}
First focus on the parts of the summand of \eqref{eq:h_ij_pairwise} that involve $Y$ (rather than $-\!\id$). Observe that $Y_{[i_0,\To,i_n,j]} = \check{R}_{j-1,j}(z_{j-1}/z_{i_n})\, Y_{[i_0,\To,i_n,j-1,j]}$. Use \eqref{eq:R-1toR'} to express this $\check{R}_{j-1,j}(z_{j-1}/z_{i_n})$ as a linear combination of $\check{R}_{j-1,j}(z_{j-1}/z_j)$ and $\id$. The terms with $\id$ arising in this way cancel against $\tilde{V}(i_0,\To,i_n,j-1,j)$ by \eqref{eq:Vtilde_cancel}. The result is proportional to 
\begin{align} \label{eq:relation}
	& \check{R}_{j-1,j}(z_{j-1}/z_j) \, Y_{[i_0,\To,i_n,j-1,j]} =  \check{R}_{j-1,j}(z_{j-1}/z_j) \\
	& \quad \times (\bigl.Y_{[i_0,\To,i_n,j-1]}\bigr|_{z_{j-1}\mapsto z_j}\! -\id) \, \check{R}_{j-1,j}(z_j/z_{j-1}) + \id \, , \nonumber
\end{align}
where the equality uses the unitarity property of the \textit{R}-matrix, $\check{R}(u)\,\check{R}(1/u)=\id$. But by virtue of \eqref{eq:Vtilde_cancel} the contribution of the final $+\id$ from \eqref{eq:relation} cancels against the remaining parts, featuring the $-\!\id$s, of the summand of \eqref{eq:h_ij_pairwise}. In this way we obtain the recursion relation
\begin{align}
	h_{[i,j]} = \ & \frac{1}{f(z_{j-1}/z_j)} \, \check{R}_{j-1,j}(z_{j-1}/z_j) \\
	& \times \bigl[f(z_j/z_{j-1})\,h_{[i,j-1]}\bigr]_{z_{j-1}\mapsto z_j} \, \check{R}_{j-1,j}(z_j/z_{j-1}) \, . \nonumber
\end{align}
By iteration this reduces to the case $i=j\,{-}\,1$. The conclusion is that $h_{[i,j]} = {-J} \, V(i-j)\,S_{[i,j]}$, as we claimed.

\section{Higher rank}

\noindent For the HSM one can replace the spin algebra~$\mathfrak{su}_2$ by $\mathfrak{su}_n$ while maintaining the model's special features~\cite{Kaw_92,*HH_92,*HH_93}. Each site carries a copy of the fundamental representation, and the Hamiltonian just involves the appropriate spin exchange. Likewise, the partially isotropic model is adapted to higher rank by using the trigonometric $\mathfrak{sl}_n$ \textit{R}-matrix. The Hilbert space is $\mathcal{H} = (\mathbb{C}^n)^{\otimes L}$. The Hamiltonian is as before, with \eqref{eq:R_matrix} generalized to (see e.g.~\cite{Jim_86})
	\begin{equation} \label{eq:R_sln}
	\begin{aligned}
	\check{R}(u) = \ & \sum_{a=1}^n E^{aa}\otimes E^{aa} + f(u) \sum_{a\neq b}^n E^{ab}\otimes E^{ba} \\
	& + g(u) \sum_{a<b}^n (u\,E^{aa}\otimes E^{bb}+ E^{bb}\otimes E^{aa}) \, ,
	\end{aligned}
	\end{equation}
with $E^{ab}$ the $n\times n$ matrix with entries $(E^{ab})_{cd}=\delta_{ac}\,\delta_{bd}$. As $q\to 1$, $\check{R}(u) \to \sum_{a,b} E^{ab}\otimes E^{ba}=P$ and we get the $SU(n)$ HSM. Since \eqref{eq:R-1toR'} and the unitarity property remain valid, the Hamiltonian may be written in Uglov's form \eqref{eq:Ham_Uglov} for higher rank too.

Analytic and numerical checks confirm that this guess works. The Hamiltonian is \textit{q}-homogeneous, and hermitian for real~$q$. It has a highly degenerate, additive spectrum, with the same \textit{q}-momentum and dispersion~\eqref{eq:dispersion}. This time, more choices of quasimomenta are allowed: $\mathfrak{sl}_n$-motifs admit at most $n-1$ adjacent $m_r$~\cite{HH+_92}. The degeneracy per motif is a bit more tricky to compute for general~$n$~\cite{BS_96,*GS_07}. The \textsc{af} motif, consisting of strings of $n\,{-}\,1$ $\tikz[baseline={([yshift=-.5*10pt*0.6]current bounding box.center)}, scale=0.2,font=\footnotesize]{
\draw[fill=black] (0,.5) circle (.5);
}$\,s separated by a $\tikz[baseline={([yshift=-.5*10pt*0.6]current bounding box.center)}, scale=0.2,font=\footnotesize]{
\draw (0,.5) circle (.5);
}$, only occurs if $n$ divides $L$. We thus have partially isotropic spin chains realizing various instances of the generalized Pauli principle. The super-case is likely likewise obtained from the $\mathfrak{sl}_{n|m}$ \textit{R}-matrix.

\section{Outlook}

\noindent Given the extensive literature on the HSM it seems reasonable to expect this work to open up new directions in the research of quantum integrability and exact solvability for long-range spin chains. Opportunities for the near future include investigating whether the interactions are sufficiently local and if the model can be adapted to the regime ${-1} < \Delta<1$ for $\Delta=(q+q^{-1})/2$ (or $q$ a root of unity), which is most relevant for realizations in nature or (cold-atom) experiments; a better understanding of the `higher-level' (affine) symmetries and the (highest-weight) wave functions; and an analysis of the thermodynamics and its field-theoretic description.

Other applications may reside in the gauge/gravity duality. The point-split form of~\eqref{eq:potential_new} and the presence of long-range multispin interactions in \eqref{eq:S_ij} resemble properties anticipated from a $\mathfrak{psu}(2,2|4)$-spin chain governing the (conformal) spectrum of $\mathcal{N}=4$ super Yang--Mills theory dual to strings moving in an $AdS_5 \times S^5$ background~\cite{Bei+_12,*Ser_11}. Although that spin chain is isotropic, a new class of examples of integrable long-range models may help finding a nonperturbative expression for it.

A particularly tantalizing direction is the study of a partially isotropic version of the Inozemtsev spin chain~\cite{Ino_03}, which should interpolate between \eqref{eq:Ham_new} and the \textsc{xxz} model~\cite{KL_18u}. The ultimate goal in this direction would be to find and solve a fully anisotropic `master spin chain' that contains the \textsc{xyz} model, Inozemtsev's elliptic spin chain and \eqref{eq:Ham_new} as special cases.

\textit{Note added.} Recently another partially isotropic HSM-like spin chain was found~\cite{SZ_18u} with pairwise interactions and no anisotropy parameter. Numerical investigations show that its spectrum is regularly spaced and highly degenerate, though the degeneracy pattern is different.

\begin{acknowledgments}
\section{Acknowledgments}
\noindent I am indebted to D.~Serban, V.~Pasquier and E.~Langmann for encouraging me to delve into \cite{Ugl_95u}. I thank N.~Beisert, J.-S.~Caux, F.~Essler, M.~Hallnäs, R.~Klabbers, H.~Rosengren, D.~Schuricht and D.~Serban for useful discussions, and F.~Finkel and A.~González-López for correspondence. I gratefully acknowledge support from the Knut and Alice Wallenberg Foundation (KAW).
\end{acknowledgments}

\bibliography{180130_references}

\end{document}